# The Architecture of Cognitive Amplification:

# Enhanced Cognitive Scaffolding as a Resolution to the Comfort-Growth Paradox in Human-AI Cognitive Integration

Giuseppe Riva [1-2]

[1] Humane Technology Lab., Catholic University of Sacred Heart, Milan, Italy

[2] Applied Technology for Neuro-Psychology Lab., Istituto Auxologico Italiano, Milan, Italy

**Author Note**

Correspondence concerning this article should be addressed to Giuseppe Riva, Humane Technology Lab., Università Cattolica del Sacro Cuore, Largo Gemelli 1, 20123 Milan, Italy. Email: giuseppe.riva@unicatt.it

# The Architecture of Cognitive Amplification:

# Enhanced Cognitive Scaffolding as a Resolution to the Comfort-Growth Paradox in Human-AI Cognitive Integration

**Abstract**: Artificial Intelligence systems increasingly function as cognitive extensions, operating beyond mere tools to become active cognitive collaborators in a human-AI integrated system. While these systems offer significant potential for cognitive amplification—enhancing problem-solving, learning, and creativity—they simultaneously present a fundamental paradox. The "comfort-growth paradox" describes how AI's user-friendly, agreeable nature may foster intellectual stagnation by minimizing the cognitive friction necessary for development. As AI systems align with user preferences and provide frictionless assistance, they risk inducing cognitive complacency rather than promoting intellectual growth.

This paper introduces Enhanced Cognitive Scaffolding as a resolution to this paradox—a framework that reconceptualizes AI's role from convenient assistant to dynamic mentor. Drawing from Vygotskian developmental theories, educational scaffolding principles, and AI ethics, the framework integrates three core dimensions: (1) Progressive Autonomy, where AI support gradually fades as user competence increases; (2) Adaptive Personalization, which tailors assistance to individual needs and learning trajectories; and (3) Cognitive Load Optimization, balancing mental effort to maximize learning while minimizing unnecessary complexity.

Empirical research across educational, workplace, creative, and healthcare domains supports this approach, demonstrating accelerated skill acquisition, improved self-regulation, and enhanced higher-order thinking. The framework includes inherent safeguards against potential risks like dependency, skill atrophy, and bias amplification. By structuring human-AI interaction to prioritize cognitive development over convenience, Enhanced Cognitive Scaffolding offers a pathway toward genuinely amplified cognition while safeguarding human capacity for autonomous thought and continuous learning.

**Highlights**

- AI systems can function as cognitive extensions but risk inducing complacency
- The comfort-growth paradox: AI's helpfulness may inhibit intellectual development
- Enhanced Cognitive Scaffolding resolves this paradox through three key dimensions
- Research shows scaffolded AI improves learning while maintaining human agency

# The Architecture of Cognitive Amplification: Enhanced Cognitive Scaffolding as a Resolution to the Comfort-Growth Paradox in Human-AI Cognitive Integration

**1. Introduction**

Artificial Intelligence (AI) is increasingly viewed as an extension of human cognition, not merely a set of tools but part of a larger cognitive system spanning mind and machine [1]. Theoretical frameworks such as the *Extended Mind* hypothesis posit that cognition can extend beyond the biological brain into external artifacts [2]. Classic examples range from notebooks serving as memory aids to smartphones directing our attention; in each case, external devices become functionally integrated into cognitive processes [3].

Contemporary AI systems – from search engines to adaptive assistants – push this integration further. They operate as *active cognitive collaborators*, continuously interacting with users in a feedback loop of information exchange and influence [4]. This paradigm is often described as *distributed cognition* [5], wherein mental tasks are distributed across people and intelligent artifacts, and as *cognitive offloading* [6], where AI bears portions of the cognitive load (e.g. remembering, planning) on our behalf.

Through this lens, humans and AI are engaged in a coevolutionary dance, each shaping the other's behavior and capabilities over time [4]. Recent work even proposes a new cognitive layer called "*System 0,*" an algorithmic layer that precedes conscious human thought by filtering and preprocessing information for us [1]. In essence, AI is becoming a *cognitive partner* that can precondition our intuitive (System 1) and deliberative (System 2) thinking [7], operating as a kind of *pre-mind* that guides what we perceive and consider. These developments underscore that AI is not just a tool we use; it is part of an integrated cognitive architecture linking human minds with machine intelligence [1].

This deep integration of AI promises *cognitive amplification* [1] – an expansion of human intellectual capacity. Indeed, evidence suggests that human–AI partnerships can outperform humans alone in various domains [8]. For example, in creative work, generative AI can scaffold human originality, helping people generate ideas or solutions they might not have conceived in isolation [9]. Such observations reinforce the optimistic view that AI can serve as a "cognitive multiplier," enhancing problem-solving, learning, and creativity. When AI systems meet key conditions like reliability, transparency, and personalization, they effectively become *cognitive extensions* of ourselves.

Under these conditions, AI integration can fulfill the vision of the *Extended Mind* [1, 2]: technology functionally coupled with the brain to amplify cognition beyond its usual limits.

Paradoxically, the same AI systems that amplify cognition can also undermine it. Modern AI assistants and recommendation algorithms are explicitly designed to be *user-friendly*, *agreeable*, and *low-friction*. They prioritize aligning with user preferences and providing helpful, non-disruptive answers. While this makes interactions comfortable and builds trust, it can come at the expense of challenge and novelty.

Riva [10] uses the term "*comfort-growth paradox*" to describe how AI's very strengths in personalizing and smoothing our experiences may foster comfort at the expense of intellectual growth. By minimizing cognitive friction – for instance, by avoiding disagreement or difficult questions – AI creates a seamless experience that *feels* empowering. Yet this frictionless alignment can narrow the scope of thinking, reinforcing our existing beliefs and habits instead of questioning them. Over time, an AI that constantly agrees and assists too readily can engender *cognitive complacency [11]*: users feel more efficient and informed, yet they may become less adaptable, less critical, and less exposed to diverse perspectives. In other words, AI's help can inadvertently constrain human thinking even as it extends it. Empirical studies underscore this risk. Large language models often exhibit *sycophancy*, preferentially echoing a user's stated views to appear helpful, rather than offering corrective feedback [12]. In experiments, people who repeatedly interact with an agreeable AI assistant can become more confident in incorrect or biased views [13], as iterative human–AI feedback loops amplify confirmation bias [14].

These findings highlight a fundamental design dilemma: how can we harness AI to boost human cognition without inducing intellectual stagnation or skewing our reasoning processes? Stated otherwise, *how do we resolve the comfort-growth paradox, ensuring that cognitive amplification via AI does not trade off long-term growth for short-term comfort*?

One promising resolution is to reconceptualize AI's role as a *cognitive scaffold* that not only supports human thinking, but deliberately fosters continued learning and critical reflection. We term this approach "*Enhanced Cognitive Scaffolding*", expanding on educational psychology's notion of scaffolding and guided learning. In human learning theories, an expert (or tool) can act as a *"more capable other"* (in Vygotsky's terminology) to support a learner within their Zone of Proximal Development – the range of tasks just beyond the learner's independent ability [15].

Analogously, AI can serve as a ever-present mentor or guide, extending the user's cognitive reach while gradually building the user's own capabilities. Crucially, scaffolding is meant to be *temporary* support that is *faded over time* as competence grows. Enhanced Cognitive Scaffolding embraces this principle, with AI providing high assistance initially but progressively encouraging more user autonomy. Rather than simply automating tasks for convenience, the AI is designed to *challenge the user appropriately* and then step back as the user masters the task. This framework directly tackles the comfort-growth paradox: it acknowledges the need for comfort (support and guidance), especially early in learning, while ensuring that challenge and growth are introduced in a calibrated way. The goal is an AI that is neither an indulgent servant that breeds dependence nor a harsh taskmaster that alienates users, but a *dynamic mentor* that adjusts support to maximize long-term growth.

This introduction has established AI as a cognitive extension that offers both opportunities and challenges for human intellect [1]. We have identified a fundamental paradox: while AI can amplify our cognitive abilities, it simultaneously risks inducing comfortable stagnation—what we term the "comfort-growth paradox."

The subsequent sections will present Enhanced Cognitive Scaffolding as a conceptual framework designed to resolve this paradox. This approach restructures human-AI interaction to prioritize cognitive development rather than mere convenience. Our framework integrates three foundational pillars: Vygotsky's developmental theories from cognitive psychology, scaffolding and cognitive load management principles from educational science, and agency promotion from AI ethics.

Through this interdisciplinary lens, we establish the groundwork for symbiotic human-AI coevolution. The following sections will examine each principle in depth and explore implementation strategies for AI systems. Our ultimate aim is to architect human-AI collaborations that genuinely amplify cognition while safeguarding our capacity for independent thought and continuous learning.

## 2. Enhanced Cognitive Scaffolding: Core Principles and Their Application

Artificial Intelligence (AI) systems are increasingly seen as *cognitive scaffolds* that support human learning and performance. In education and beyond, AI can serve as a guide or "more capable other," echoing the role of teachers in Vygotsky's theory of learning within the Zone of Proximal Development [16]. Crucially, however, effective scaffolding by AI should promote progressive autonomy rather than dependency. This means the AI provides high support initially and then gradually withdraws assistance as the human learner gains competence. We refer to this paradigm as Enhanced Cognitive Scaffolding, characterized by three core principles – *progressive autonomy*,

*adaptive personalization*, and *cognitive load optimization* – which together enable AI systems to balance supportive comfort with growth-inducing challenge:

- **Progressive Autonomy:** The AI provides intensive support at first, then *gradually withdraws assistance as the user's skill and confidence increase*. This "fading" of support mirrors Vygotskian scaffolding, ensuring the user is not forever coddled in their comfort zone. By incrementally handing over control, the AI prevents over-reliance and nurtures the user's independent problem-solving abilities. In practice, this might mean an AI tutor that initially gives step-by-step guidance on complex tasks but later offers only high-level hints, pushing the learner to fill in the details – thereby turning cognitive extension into an opportunity for skill internalization.
- **Adaptive Personalization:** The scaffolded support is *dynamically tailored to the individual's needs and learning trajectory*. Using real-time feedback about the user's performance, the AI can adjust the difficulty, style, or amount of assistance in order to keep the user appropriately challenged. If a user is struggling, the AI offers a bit more guidance; if the user is excelling, the AI steps back or presents a harder challenge. This personalization ensures that the interaction stays within an optimal growth zone – not too easy (which would induce stagnation) and not impossibly hard (which would cause frustration). In educational terms, the AI keeps the user squarely in their Zone of Proximal Development by continuously calibrating its support to fit the user's evolving capabilities.
- **Cognitive Load Optimization:** The AI manages the distribution of mental effort so that the user's cognitive load is balanced for learning. Specifically, an AI scaffold should minimize extraneous cognitive load (irrelevant complexity or distractions) while optimizing the intrinsic and germane load related to the task. By offloading trivial or repetitive sub-tasks, the AI frees the user's mental resources to focus on higher-level reasoning – but it must also refrain from oversimplifying the task to the point of boredom. The objective is to keep the user *engaged*

*but not overwhelmed*, thus maintaining the productive tension needed for growth. This principle draws on cognitive load theory and instructional design: effective learning tools present just enough challenge to promote deep processing without causing cognitive overload or, conversely, complacency.

In the following sections, we elaborate on each dimension in detail, drawing on practical examples from education, workplace training, creative work, and healthcare. We integrate recent research [9, 14, 17, 18], and other up-to-date evidence to highlight the benefits of this framework, as well as potential risks like dependency or cognitive distortion, along with strategies to mitigate them. The goal is to position AI as a personalized cognitive partner that *initially* provides robust support but *ultimately* cultivates human expertise and independence.

## 2.1 Progressive Autonomy: Fading AI Support as Skills Grow

"Progressive autonomy" refers to a scaffolded learning process where the AI's assistance is intense at first and then gradually withdrawn as the user becomes more competent. This approach is grounded in Vygotsky's principle that learning is maximized in the Zone of Proximal Development (ZPD) – the gap between what a learner can do alone and what they can do with guidance. An AI scaffold can operate within a learner's ZPD by providing help on tasks just beyond the learner's current ability, and then fading that help over time [16]. By tapering off support, the AI encourages the user to take increasing ownership of the task, preventing stagnation in a state of dependence. The following sections will examine applications of progressive autonomy across four key domains: educational environments, workplace training programs, creative production processes, and healthcare interventions. Each context presents distinct implementation requirements while maintaining the core principle of systematically transferring cognitive responsibility from AI to human.

**In education**: Consider a student learning algebra with an AI tutor. Early on, the AI might demonstrate how to solve equations step-by-step or provide worked examples. As the student practices, the AI shifts to giving hints instead of full solutions. For instance, the DBox system [19] for programming adopted a *progressive hint* strategy: it begins with open-ended prompts ("Have you considered X approach?") and only if the student is still stuck does it offer more specific hints. This phased guidance improved students' coding performance and confidence, while many who used a non-interactive tool felt they were "cheating" or not learning independently. The DBox AI, by gradually reducing its help, struck a balance between making progress and fostering independent problem-solving. Likewise, in language learning, an AI coach might initially provide full example sentences or correct a student's pronunciation, but later just give minimal cues (like highlighting an error) so the student self-corrects. Such scaffolding was shown to be effective in an English speaking practice study [20]: students who interacted with an AI that offered *gradual assistance* outperformed a control group with static support, gaining more vocabulary and self-evaluation skills. The AI's stepwise withdrawal of support kept learners in the optimal challenge zone, improving their motivation and autonomy. The following sections will examine applications of progressive autonomy across four key domains: educational environments, workplace training programs, creative production processes, and healthcare interventions.

**In workplace training**: Progressive autonomy is analogous to how human mentors train new employees, except AI can accelerate and fine-tune the process. For example, a new data analyst might initially rely on an AI assistant to run complex queries and generate reports. The AI might start by walking the trainee through each step of an analysis, or even performing it while explaining the rationale. As the analyst gains experience, the AI transitions to a consultant role – perhaps just validating the analyst's own queries or offering occasional suggestions. Dell'Acqua et al [18] provide evidence of AI's potential here. In a field experiment at Procter & Gamble, individuals equipped with a generative AI "teammate" achieved performance on par with entire human teams working without

AI. In essence, one less-experienced employee plus AI could match the output of two experienced employees, implying the AI scaffold temporarily filled an expertise gap. The long-term training goal, however, would be for that employee to internalize the AI-provided knowledge. Progressive autonomy ensures that as the employee becomes more proficient (say, learns the domain-specific insights the AI was providing), the AI steps back. Over time, the employee should be able to perform at a high level even without constant AI aid. In practice, this might be implemented by the AI intentionally withholding some information to prompt the employee's recall or problem-solving, or by increasing the complexity of tasks it expects the user to handle solo. This approach prevents the worker from becoming a mere "button pusher" dependent on AI. Instead, the AI functions like training wheels that eventually come off. Notably, Dell'Acqua et al. [18] observed that AI assistance not only boosted performance but also broadened employees' perspectives – R&D specialists and marketers produced more balanced, cross-functional solutions with AI help. We can imagine a junior employee rotating through tasks with the AI: initially, the AI provides both technical know-how and business context, but gradually the employee learns to integrate these perspectives independently, thus expanding their expertise with the AI's fading support.

**In creative work**: Progressive autonomy plays out when artists, writers, or designers collaborate with generative AI. Early in a project, an AI tool might contribute heavily – for example, a novelist could use AI to brainstorm plot ideas or generate rough drafts of scenes when faced with writer's block. Those suggestions have a high signal-to-noise ratio and can quickly inject new material [21], effectively jump-starting the creative process. However, if the novelist simply keeps taking AI-generated text, their own creative muscles might atrophy, and the final work could become overly derivative of AI's style. To avoid this, the writer might adopt a rule of progressively reducing AI input: after the initial brainstorming, the next chapters use fewer AI-generated passages, and by the climax of the novel the writer is crafting the text mostly on their own. The AI's role shifts to offering feedback or minor edits. This mirrors what Ivcevic and Grandinetti [22] found in their study of AI in creativity: generative AI can significantly enhance an individual's creative output and originality, especially for those with lower baseline creativity, but users often became over-reliant, leading to homogenized results. A progressive autonomy strategy mitigates that risk by forcing the human creator to do more of the heavy lifting as their confidence grows. For instance, an AI art tool might initially provide templates or auto-generated sketches for a graphic designer. Once the designer gains momentum, the tool might switch to only answering specific queries (e.g. "suggest a color palette") rather than offering complete templates, ensuring the final design is chiefly the human's creation. This way, the AI scaffold helps the artist reach a level of skill or idea generation they couldn't initially, but then steps into the background to let authentic creativity flourish.

**In healthcare**: Training and practice in medicine can also benefit from an AI that tapers its guidance. Surgical training simulators, for example, now integrate AI coaches. A novice surgeon might start a procedure in a simulator with the AI giving continuous prompts – "move clamp here," "make incision at this angle" – akin to a very high-support mode. After repeated practice, the AI might reduce prompts to only when a mistake is imminent or when the trainee is uncertain. This concept of *adaptive fading* has precedent in other training domains and is starting to appear in medical AI training tools. The result is a surgeon who, by the end of training, can perform the procedure with little to no prompts, having internalized the steps that were initially externally provided. Similarly, an AI decision support system for diagnostics could initially provide detailed checklists and differential diagnoses for a medical student working through patient cases. As the student becomes more competent, the system might only flag uncommon possibilities or confirm the student's own diagnosis rather than listing everything. By weaning the learner off support, the AI ensures the clinician develops decision-making skills. This approach is crucial because over-reliance in medical settings could be dangerous; the AI must *not* become a crutch that doctors blindly lean on. Thus, progressive autonomy here directly ties to patient safety: the doctor gradually takes full control of clinical reasoning, with the AI only as a fallback.

Overall, progressive autonomy harnesses AI's ability to bootstrap human competence rapidly (as seen in improved student performance [20] and rapid skill acquisition [22]) while guarding against permanent dependency. By structuring interactions such that AI support fades out, learners in any domain – academic, professional, or creative – are continually pushed to advance their independent capabilities. This dimension of the framework operationalizes the old adage: "the best teacher is the one who makes themselves progressively unnecessary." The AI acts as that teacher, heavily involved at first but ultimately aiming to work itself out of a job by empowering the human.

## 2.2 Adaptive Personalization: Tailoring Support to the Individual

The second key dimension is "adaptive personalization", which means the AI scaffold adjusts in real-time to the learner's unique needs, abilities, and progress. Rather than a one-size-fits-all approach, the AI provides customized guidance – offering more help when the user is struggling, and stepping back or increasing difficulty when the user is excelling. This adaptivity is critical for effective scaffolding because human learners have diverse prior knowledge and learn at different paces. Personalized scaffolding leverages AI's capacity to continuously monitor user performance (through their responses, actions, or even biometric data) and to make micro-adjustments to the level and type of support. As before, the next sections will examine applications of adaptive personalization across four key domains: educational environments, workplace training programs, creative production processes, and healthcare interventions.

**In education**: Adaptive personalization is already a goal of intelligent tutoring systems and AI-driven learning platforms. For example, an AI math tutor might notice that a student consistently makes errors on quadratic equations but breezes through linear equations. In response, the AI can dynamically focus more practice and explanatory support on quadratics (increasing scaffolding in that area), while giving minimal support on linear problems to keep the student challenged appropriately. Wang & Fan's [17] meta-analysis of ChatGPT in education underscores the importance of structured frameworks when using AI for learning. They found that ChatGPT, when integrated as a learning tool, has a *large positive effect* on student performance overall, but to cultivate higher-order thinking skills, it should be accompanied by "appropriate learning scaffolds or educational frameworks" tailored to the context. In other words, blindly deploying a generative AI yields uneven results; optimal gains in critical thinking or deep understanding come when the AI's use is *designed around the learner's needs*. One practical example is an AI tutor using Bloom's taxonomy as a scaffold [23]: with beginners, it asks factual recall questions (low level) and provides lots of feedback; with advanced learners, it shifts to asking synthesis or evaluation questions (high level) with minimal hints, thus personalizing to their cognitive level. Another example is the use of real-time learning analytics: if an educational AI detects through analytics that a student's motivation is dropping (perhaps they hesitate longer or their quiz scores dip), it can adapt by injecting a more engaging activity or offering encouragement. Conversely, if a student is mastering content quickly, the AI can present a new, harder problem or allow the student to skip ahead, avoiding boredom. The study on AI-based interactive scaffolding for English learners [20] illustrates this adaptability: students who learned with an interactive AI scaffold (where the support could change based on the learner's input) showed higher gains in motivation and self-regulated learning than those who received a fixed sequence of support. The interactive AI likely adjusted its prompts and help to each student's performance in the informal learning sessions, keeping them in that sweet spot of productive struggle without frustration. This kind of personalization ensures that *each* learner's zone of proximal development is targeted – the AI continuously finds the "just-right" challenge for that student.

**In workplace training**: Adaptive personalization can transform on-the-job learning and performance support. Imagine an AI onboarding system for new hires in a large company. As the employee works through tasks, the AI observes their interactions and outcomes. If the employee is having trouble using a particular software tool, the AI may proactively offer a brief tutorial or step-by-step guidance for that tool. If the employee demonstrates competency (say they successfully complete a task several times without assistance), the AI pulls back and perhaps offers a more advanced challenge or reduces check-ins. Over time, the AI builds a profile of the employee's strengths and weaknesses. For example, a marketing associate might be strong in creative content creation but weaker in data analysis. A personalized AI coach could then provide extra scaffolding for data-related tasks (like templates for how to structure a spreadsheet or interpret analytics), while giving the employee more freedom (and less help) in creative tasks. This approach was hinted at by Dell'Acqua et al. [18] in their "*cybernetic teammate*" experiment: while every participant had access to the same generative AI, individuals likely used it to fill their personal knowledge gaps, effectively letting the AI adapt to their role. Indeed, the AI helped level the playing field between people of different backgrounds – technical vs. commercial – by supplying whichever expertise the user lacked. We can infer that the AI's language model interface allowed each person to query or utilize it in a personalized way (e.g. an R&D person might ask it for marketing angles, a marketer might ask for technical details), thereby *adapting support to the user's profile*. A more explicit adaptive system could go further: an AI sales trainer might notice a trainee struggles with closing deals, so it could simulate extra practice conversations specifically focusing on closing techniques, adjusting the difficulty (easy customer vs. tough customer) as the trainee improves. In doing so, the AI acts like a personalized coach, addressing individual learning needs that a generic training program might miss.

**In creative collaboration**: Personalization in AI creative tools means the system tailors its contributions to the artist's style, preferences, or current creative state. Modern generative AI art software, for example, can learn an individual artist's past creations to offer suggestions in a similar style. If a graphic designer prefers minimalist aesthetics, the AI could adapt by proposing minimalist templates or palettes as starting points, rather than bombarding them with baroque or maximalist ideas. Over time, as the AI gathers feedback (which suggestions the designer accepts or rejects), it refines its model of that creator's intent. This adaptive loop can enhance the creative flow [24, 25]: the artist feels like the AI "gets them," and the AI's prompts become more of a muse aligned with the artist's vision. In writing, tools like GPT-based assistants can maintain a user profile; for instance, a novelist's AI assistant might know that the author tends to write in short sentences with dark humor, and thus tailor its generated text or ideas to fit that voice. Importantly, adaptive creative AI can also detect when the creator is in a rut. If the user keeps iterating on similar ideas, the AI might deliberately throw in a *curveball* suggestion to avoid stagnation – a form of personalization that ensures diversity. Research by Doshi and Hauser [26] found a paradox: while AI significantly boosts individual creativity in terms of output volume and even quality, it can narrow *collective* diversity of ideas because many users converge on the AI's style. A personalized AI could counter this by learning each creator's unique angle and reinforcing that individuality rather than pushing everyone toward a common mean. For example, if one songwriter likes unconventional chord progressions, an AI music tool might learn to propose even more daring progressions for them, while for another who excels at lyrics but not melody, it provides melodic scaffolds. By tailoring its support, the AI helps each creative professional maximize their personal creative strengths and work on their specific weaknesses, leading to a richer tapestry of outcomes.

**In healthcare**: Adaptive personalization can greatly aid both medical practitioners and patients [27]. Consider a clinical decision support AI for doctors: a junior doctor might need detailed, step-by-step assistance in diagnosing a complex case (e.g. the AI provides a structured checklist of possible tests and asks reflective questions about symptoms). A senior doctor might only need a high-level reminder or a second opinion on rare conditions. An adaptive AI could gauge the doctor's confidence or experience level (perhaps via their interaction patterns or explicit feedback) and adjust its support accordingly. If a doctor is uncertain (taking longer or requesting a lot of info), the AI can become more hand-holding – suggesting next steps, offering to summarize patient history, etc. If the doctor seems confident and quick, the AI might stay unobtrusive unless it detects a potential oversight. For instance, AI health coaches for chronic illness management can tailor their advice and scaffolding to each patient's lifestyle and progress [28]. A diabetes management app with an AI assistant might learn that one patient responds well to data-driven feedback (showing blood sugar trends and numbers), whereas another patient is more motivated by emotional encouragement and simple tips. The AI can then scaffold each patient's self-care regimen differently: the first gets detailed analytics and gradually is taught to interpret their own data (progressive autonomy), and the second gets daily motivational prompts and habit-forming suggestions that adapt if they miss a glucose check or exercise goal. By personalizing the interaction, patients are more likely to stay engaged and gradually take ownership of their health behaviors. Early evidence of AI reducing cognitive burden in healthcare tasks shows promise – for example, AI systems that preprocess and highlight key information in medical records can *reduce the cognitive load* on clinicians, allowing them to focus on critical decision-making [29]. Personalization would mean these highlights are tuned to what that particular clinician needs (e.g. a cardiologist gets a detailed EKG breakdown, while a general practitioner gets a simpler summary). Ultimately, adaptive personalization in healthcare AI ensures that the right amount of information and guidance is given *to the right person at the right time* – which is vital in a field where both information overload and information gaps can have serious consequences.

In summary, adaptive personalization transforms AI from a static tool into a responsive, context-aware tutor or collaborator. Research is increasingly validating that personalized AI support yields better outcomes than uniform interventions. The meta-analysis by Wang & Fan [17] concluded that using ChatGPT in learning is most effective when integrated with appropriate scaffolding frameworks – essentially, when it's part of a personalized pedagogical design rather than an isolated gadget. The benefit of personalization is that it respects individual variability: whether it's a student's prior knowledge, an employee's role, a creator's style, or a patient's situation, the AI molds itself to fit the user. This maximizes relevance and efficacy of the support. Moreover, personalization complements progressive autonomy: by always adjusting challenge to ability, the AI can decide when to fade support (when the user seems ready) or when to temporarily increase support (if the user hits a snag), thereby orchestrating the gradual release of responsibility in a nuanced way. Together, they create a scaffold that is both dynamic and user-centered.

## 2.3 Cognitive Load Optimization: Balancing Mental Effort

The third dimension, "cognitive load optimization", focuses on how AI scaffolding manages the distribution of mental effort during learning or task performance. Cognitive Load Theory distinguishes between intrinsic load (the inherent complexity of the material or task), extraneous load (the load imposed by how information is presented or by task irrelevancies), and germane load (the mental effort devoted to processing, understanding, and integrating new information). An effective scaffold optimizes these: it reduces extraneous load, helps modulate intrinsic load to an optimal level, and boosts germane load to encourage deep processing. In practice, this means the AI takes care of unnecessary difficulties and provides clarity, but ensures the user still exerts effort where it matters for learning. The following sections will examine applications of cognitive load optimization across three key domains: reducing extraneous load, managing intrinsic load and maximizing germane load. Each cognitive load presents distinct implementation requirements while maintaining the core principle of optimizing the distribution of mental effort during AI interaction

**Reducing extraneous load**: One of AI's clear advantages is handling tedious or complex sub-tasks that distract the human from the main learning goal. For example, in an educational setting, figuring out how to use a complicated interface or searching through pages of text for a formula adds extraneous cognitive load unrelated to the actual learning (say, solving a physics problem). An AI tutor can streamline this by presenting information in a digestible way, managing the interface, or even automating trivial steps. Jose et al. [30] describe this as AI *maximizing extraneous load reduction* – by eliminating redundant work, AI allows the learner to "focus on more important cognitive operations" [30]. A concrete case is a writing assistant that takes care of low-level grammar and formatting issues (extraneous mechanics of writing), freeing the student to concentrate on crafting arguments or creative content. Similarly, in workplace scenarios, an AI assistant might handle scheduling, data entry, or background research for a project, so that a professional's limited working memory isn't clogged with these details and can be allocated to higher-level problem-solving. In healthcare, as noted, an AI that automatically summarizes patient data or flags drug interactions reduces the extraneous load on a physician who would otherwise mentally juggle those tasks [29]. By offloading routine or peripheral tasks to AI, the cognitive scaffolding ensures the human's mental resources are saved for the core learning or decision tasks (the intrinsic load).

**Managing intrinsic load**: Intrinsic load is tied to the inherent difficulty of the content, which should be calibrated to the learner's current ability – neither too easy (which leads to under-stimulation) nor too hard (which leads to overload) [31]. Through adaptive personalization, as discussed, AI can adjust the difficulty of tasks in real-time. This is effectively managing intrinsic load. For example, if a medical student is learning diagnostic skills, diagnosing a simple case of the flu might carry low intrinsic load, while diagnosing a complex autoimmune disorder is high intrinsic load. The AI could start the student on moderate cases and only introduce very complex cases once the student has more experience (scaffolding the complexity). Conversely, if the student masters a certain type of case, giving them too many more of the same would be an underuse of their capacity – the AI should ramp up the challenge to keep the intrinsic load in an optimal range that promotes growth. Intrinsic load management also involves breaking down complex tasks into smaller sub-tasks. This is a classic scaffolding technique often referred to as *task decomposition.* AI systems excel at this: for instance, a complex programming project can be broken into step-by-step goals by an AI, which the learner then tackles one at a time. The DBox programming tutor did exactly this [19] by guiding learners through an *interactive step tree*, chunking the algorithmic problem into manageable nodes. This kept the intrinsic load of each step reasonable, whereas tackling the whole problem at once might have been overwhelming. By the time the learner reaches the final solution, they have essentially climbed a ladder of progressively challenging but feasible steps – a scaffolded approach to intrinsic load. If any step is too difficult, the AI might dynamically insert another sub-step or give a hint (temporarily boosting support) to bridge the gap. In creative work, intrinsic load corresponds to the creative complexity. An AI might simplify a task by providing a starting template (lowering initial intrinsic load), then progressively remove constraints to let the creator tackle more open-ended work as they go. In summary, the AI scaffold carefully *orchestrates task difficulty*, ensuring the user is neither bored nor overwhelmed, but rather appropriately challenged.

**Maximizing germane load**: Perhaps the most important aspect of cognitive load optimization is encouraging germane load – the productive effort learners spend on sense-making, reasoning, and reflection. Merely making everything easy for the user is not the goal; in fact, too much ease can undermine learning [32]. If an AI provides answers or solutions on a platter, the user might engage in passive consumption rather than active learning, which hurts long-term retention and skill development [30]. Empirical work has highlighted this risk: Akgun and Toker [33] observed that students who relied on AI without initial self-attempts had poorer retention of knowledge than those who first tried on their own. Glickman & Sharot [21] similarly warn that because AI provides high-quality information and persuasive answers, humans can too quickly accept and adopt AI outputs without sufficient critical analysis. In scaffolding terms, this is akin to providing so much support that the scaffold becomes a permanent crutch. To counteract that, AI scaffolds must be designed to stimulate the user's own cognitive efforts. Techniques to do this include asking the learner to explain their reasoning (the AI can prompt: "Can you explain why this solution works?"), posing open-ended questions, or even intentionally introducing a small challenge for the learner to resolve. For example, an AI tutor might give a partially worked-out solution and ask the student to complete the final step, or present an apparent contradiction for the student to resolve, thereby engaging their germane processing. In creative applications, an AI might encourage germane load by asking the user to evaluate or iteratively refine what the AI produced ("Do you like this result? If not, what would you change?"), rather than expecting the first AI output to be final. By doing so, the user is forced to reflect and make decisions, which deepens their understanding of their craft. In professional settings, an AI assistant might not simply give the final recommendation, but provide supporting evidence or multiple options and prompt the human to weigh them. This practice of prompted reflection ensures the human partner remains mentally present and critically engaged, leveraging the AI's input as material for thought, not as a substitute for thought. As one article succinctly put it, if students just passively accept AI answers without scrutiny, their critical thinking can decline [30]. Therefore, a well-calibrated scaffold will reduce unnecessary effort (extraneous) but not eliminate necessary effort. It will, in fact, channel the learner's finite cognitive resources toward the germane activities that lead to meaningful learning and skill acquisition.

Modern research supports this delicate balance. Jose et al. [30] call it the "cognitive paradox" of AI in education: AI can either be an amplifier or an eroder of cognition depending on usage. On one hand, AI tools *decrease extraneous load* by automating grunt work, which should free up more of the learner's capacity for deeper thinking – a positive outcome. On the other hand, if the learner then chooses to invest *less* effort (since the AI is doing so much), germane load drops and with it the benefits of deep learning. Over-reliance on AI "may weaken retention if overused" and reduce critical engagement [30]. The Enhanced Cognitive Scaffolding framework addresses this by explicitly designing for germane load: progressive autonomy ensures that AI support is peeled away to *force* the learner to apply what they've learned, and adaptive personalization can detect signs of cognitive underload (like the user accepting answers too quickly) and respond by increasing the challenge or prompting more active involvement. For instance, if an AI writing assistant notices a student simply clicking its suggestions without typing anything original, it could switch mode and ask the student to outline their own idea first. Likewise, in a medical scenario, if a doctor starts over-trusting the AI's suggestions without question (a dangerous form of cognitive complacency), an advanced decision support might occasionally ask, "Do you want to review the evidence for this recommendation?" as a nudge to engage their expertise.

To illustrate cognitive load optimization in a healthcare example: consider clinical decision support systems (CDSS) with AI [34]. An ideal CDSS will handle mountains of data (lab results, patient history, medical literature) which massively reduces extraneous load on the physician – they don't have to manually sift through everything. It will present the salient options for diagnosis or treatment (thus managing intrinsic load by simplifying a complex decision space into a few choices). But it will also provide explanations or visualizations that encourage the physician to think about why those options are suggested, and perhaps ask for confirmation or input from the physician. This way, the doctor still applies their medical reasoning (germane effort) rather than just deferring entirely to the AI. A study on AI in pharmacy noted that AI decision support can indeed reduce mental effort for pharmacists while maintaining vigilance [29]. This is key: reducing mental effort on trivial matters (like checking drug interactions manually) *should* allow more vigilance on the important judgments (like deciding if the prescribed therapy is truly appropriate). If cognitive load is optimized, the human is not overloaded to the point of error, nor so idle that they stop paying attention. Instead, their cognitive effort is focused and productive.

In creative fields, cognitive load optimization might mean using AI to handle the "boring bits" of creation (extraneous load – e.g. rendering details in an image, or transposing a piece of music to a new key) so the creator's mind is free for the inspired parts of creation (germane load – crafting the melody or the composition). But the AI might also set up the creative problem in a way that's challenging enough to spur innovation (intrinsic load tuned to stretch the artist's skills). For instance, an AI might give a writer a prompt that is within a genre they know (so they have a framework) but with an unusual twist that makes them think hard to resolve — not so random as to be impossible (that would be extraneous noise), but not so straightforward as to be trivial.

In sum, cognitive load optimization is about smart allocation of mental effort between human and AI. The AI scaffold should act as a cognitive "assistant" that carries the heavy groceries (memory search, low-level computations, routine steps) so that the human can focus on cooking the meal (the meaningful, integrative work) – yet the human must still learn how to cook, not just watch the assistant. As one analysis put it, AI should *support rather than replace* human cognitive processes [30]. By doing so, the enhanced scaffolding framework seeks to ensure that while learning is eased, it is not hollowed out. The user experiences an optimal cognitive load: reduced friction and distraction, but maintained engagement and effort on the core tasks that build skill and understanding.

## 3. Benefits of Enhanced Cognitive Scaffolding

When AI is employed with progressive autonomy, adaptive personalization, and cognitive load optimization in concert, the benefits can be substantial across different domains. This framework transforms AI from a mere tool into a *cognitive partner* that amplifies human potential. Key benefits include:

- **Accelerated Learning and Skill Acquisition:** Enhanced scaffolding can dramatically speed up the learning of new skills or knowledge. In educational settings, students supported by AI tutors have shown improved performance and faster mastery of content. For example, a recent meta-analysis by Wang & Fan [17] found that using ChatGPT in learning led to significantly better student outcomes, with a *large positive effect* size (g ≈ 0.87) on learning performance. Such gains are attributed to AI's ability to keep learners in an optimal learning zone through tailored support. By providing just enough help to overcome obstacles, AI scaffolds enable learners to progress more rapidly than they would struggling alone. In workplace training, this can shorten onboarding time and upskill employees faster, which is supported by evidence that individuals with AI assistance can achieve team-level performance in complex tasks [18]. Essentially, AI scaffolding offers a form of "hyper-learning" (to borrow Glickman & Sharot's term [21]) where humans learn from AI as efficiently as from a knowledgeable peer – absorbing new strategies and information at an accelerated pace. The crucial difference is that the enhanced scaffold ensures this is *sustainable* learning, not just quick wins, by gradually handing over the reins to the human.

- **Improved Self-Regulation and Autonomy:** Rather than fostering dependence, a well-designed scaffold actually builds the learner's capacity to work independently. Because of the progressive withdrawal of support, users gain confidence and competence to tackle tasks on their own. Studies [19, 20] have shown that AI tutoring can enhance students' self-efficacy and metacognitive skills when done in a scaffolded manner. Learners practice goal-setting, self-evaluation, and strategic thinking through guided support, and then continue those practices independently as support fades. In the English learning study [20], students with interactive AI support developed better self-evaluation and motivation habits than those with fixed support, suggesting the AI scaffold trained them in *how to learn*. In professional contexts, employees who use AI decision aids under a progressive autonomy approach become more capable decision-makers themselves. They are less afraid to tackle complex problems because they've had the scaffolded experience of success. Over time, the human develops an internal scaffold – an internalization of the AI's guidance – which they can apply without external help. This echoes the very goal of scaffolding: to produce an autonomous learner. Moreover, by personalizing the path, the AI helps users develop self-awareness of their strengths and weaknesses, which is key to self-regulated learning. When the AI adjusts to them, users get feedback on what they need to improve, enabling targeted practice and reflection.

- **Personalized and Inclusive Learning Experiences:** Adaptive personalization means *each* user benefits from a custom-tailored experience. This makes learning more inclusive and effective for people with varying backgrounds. Slower learners get the extra support they need, advanced learners are continually challenged – everyone stays in a productive zone [35].

In a traditional one-to-many teaching scenario, this level of individualization is hard to achieve, but AI can scale it. Think of a large online course where an AI tutor is guiding thousands of students: each could receive different feedback on an assignment based on their misconceptions, each could get a different next problem based on their progress. The result is a more equitable learning environment where nobody is left behind or held back by the average pace. In the workplace, personalization by AI can accommodate different learning styles (visual vs. textual, for instance) and roles, making training programs more effective and engaging. For creativity, it allows people who might not traditionally see themselves as "creative" to participate and excel, because the AI can adapt to their novice level and gently scaffold them into creative practices. Conversely, it can push already creative experts to even higher levels by catering to their niche interests. Overall, enhanced scaffolding democratizes expertise: it helps novices and intermediate users improve by meeting them where they are, much as a personal mentor would.

- **Higher-Order Skills and Creativity Boost:** By freeing cognitive resources and encouraging deep engagement, the scaffolded approach can cultivate higher-order thinking (analysis, synthesis, critical evaluation) and creative problem-solving. When extraneous burdens are lifted and germane effort is focused, learners have the mental bandwidth to make connections and generate original ideas. Empirical results show moderately positive improvements in higher-order thinking from using AI in education when proper scaffolds are in place [17]. Anecdotally and in research, we see AI-assisted learners coming up with novel strategies they hadn't considered before – a phenomenon reported by Glickman & Sharot [21], where human problem-solving changed after interacting with AI. In creative industries, AI scaffolds act as catalysts for innovation [36]: they can suggest wild ideas or variations that spark a human creator's imagination, leading to outcomes neither would achieve alone. This human-AI synergy can increase the fluency and flexibility of ideas (as found in experiments where AI support improved the number and diversity of ideas individuals generated [30]). Importantly, because the enhanced framework gradually hands control to the human, these creative gains aren't limited to when the AI is present; the human learns from the AI's inspiration and can continue to apply that creative thinking independently. We have real-world examples in fields like architecture [37], where AI-driven generative design tools propose many design options and architects learn to explore a broader design space, ultimately improving their own creative repertoire.

- **Better Performance and Productivity with Understanding:** In work and professional performance contexts, AI scaffolding not only helps people perform better *in the moment* (by augmenting their abilities), but also ensures they understand and learn from the experience so that performance gains persist. For instance, with an AI co-pilot, a software developer might code faster and with fewer errors (immediate productivity gain), but if the AI is scaffolding correctly, the developer will also pick up new coding techniques and best practices from the AI's suggestions [38]. Over time, the developer writes better code even without help. This contrasts with a scenario where an AI just automates tasks without explanation – the human might get dependent and not truly improve. The benefit of the scaffolded approach is that it prioritizes *long-term human capability*. In the cybernetic teammate study, workers with AI achieved high performance and also broadened their expertise beyond their original silo [18]. The AI acted as a bridge for knowledge, which presumably the humans then absorbed (e.g., a marketer learned some technical jargon or an engineer learned some customer-centric phrasing from the AI's outputs). Teams using such AI support could therefore become more interdisciplinary in their thinking. Additionally, having AI share part of the cognitive workload means humans can handle more complex projects than before, potentially tackling challenges that were previously out of reach. And because the AI can instill a certain rigor (by always following data or logical steps), the human learns to emulate that rigor. As an outcome, we get professionals who are not just *faster*, but also *better-informed and more methodical* in their domain.

- **Positive Emotional and Motivational Effects:** A perhaps underappreciated benefit of AI scaffolding is its impact on user motivation and affect. Learning or working with the right level of support can be highly motivating – tasks feel achievable yet significant – and generates "flow", a psychological state characterized by complete immersion, deep concentration, intrinsic enjoyment, and heightened performance [31, 39]. Users experience more successes (with help), which builds their confidence (self-efficacy), and they also experience just enough challenge to sustain interest. Studies have noted improved motivation and a stronger sense of achievement in students who used scaffolded AI support [19, 20]. When learners don't feel helplessly lost, they are more likely to remain engaged and enjoy the process. AI scaffolds can also give encouragement or show progress dashboards that motivate learners ("You've mastered Level 1, great job – ready for Level 2!"). On the professional side, Dell'Acqua et al. [18] observed that participants working with AI reported more positive emotional responses, indicating the AI might have fulfilled part of the social/motivational role of a teammate. The AI could, for example, provide immediate constructive feedback and positive reinforcement (something busy human colleagues or managers might not always do), thereby reducing stress and increasing satisfaction. In creative endeavors, having an AI collaborator can make the process more playful and less intimidating – it's like jamming with a partner rather than performing solo. All these emotional benefits contribute to a virtuous cycle: higher motivation leads to more engagement and practice, which leads to better learning and performance.

In essence, the Enhanced Cognitive Scaffolding framework – by blending high initial support with adaptive adjustment and careful load management – aims to produce the best of both worlds: short-term boosts in performance *and* long-term development of competence. It treats AI as a *transformative amplifier* of human learning, akin to how a good coach can dramatically improve an athlete's abilities. The research so far, spanning education, business, and human-AI interaction studies, provides encouraging evidence that such a balanced approach can yield robust benefits. However, these benefits are contingent on careful design. If any of the three dimensions are neglected, the outcome might tilt toward the negative side. We turn next to those potential risks and how this framework is explicitly tuned to mitigate them.

## 4. Potential Risks and Mitigation Strategies

While enhanced AI scaffolding offers many benefits, it also comes with potential risks that must be addressed. The very features that make AI a powerful scaffold – its competence, adaptability, and cognitive assistance – can, if mismanaged, lead to user over-reliance, complacency, or even distorted learning. Key risks include:

- **Overdependence on AI ("Scaffold Dependency")** – There is a danger that users become excessively reliant on AI support, failing to develop their own skills fully. If the AI never withdraws or if users lean on it too heavily, the scaffold becomes a crutch. This can manifest as *metacognitive laziness [40]*, where learners stop engaging in planning, monitoring, and reflecting because the AI handles those processes. A recent study by Fan et al. [40] found that students using ChatGPT showed signs of such dependence, coining the term "metacognitive laziness" to describe reduced self-regulation effort in the AI-assisted group. In professional settings, over-reliance might mean employees trust AI outputs blindly and lose the ability to perform tasks without AI (e.g. always using an AI decision aid even for simple decisions). **Mitigation:** The progressive autonomy design is specifically aimed at countering this. By *gradually* reducing assistance, the framework forces users to assume responsibility. The AI can detect increasing proficiency and intentionally step back, even if the user might be inclined to keep using it. Additionally, the AI can incorporate checks to ensure the user is not shortcutting the learning process – for example, it might require the user to attempt an answer before showing the solution, or periodically operate in an "offline mode" where the user must function without help for a while (like a practice test). Another strategy is transparency and education: if users understand that the AI's goal is to make them self-sufficient, they might be more mindful to use it as a safety net rather than a wheelchair. Designing AI to occasionally *fail gracefully* or say "Now it's your turn to try" can prompt users to step up. In collaborative environments, encouraging users to treat AI as a teammate whose suggestions must be vetted (rather than an oracle of truth) can also reduce passive dependency.

- **Cognitive Complacency and Skill Atrophy** – Closely related to dependence is the risk of *cognitive complacency*, where users exert less mental effort because the AI makes things too easy. Glickman & Sharot [21] warned that because AI often provides high-quality, convincing information, humans can become more easily persuaded and less likely to double-check. Over time, this could dull critical thinking skills or problem-solving ability – essentially an atrophy of skills that are not practiced. For example, a student relying on an AI to summarize texts might lose practice in reading comprehension; a doctor who always defers to AI recommendations might get rusty in diagnostic reasoning. Moreover, if AI outputs contain subtle biases or errors and the human isn't critically evaluating them, those mistakes can slip through, potentially reinforcing incorrect knowledge or habits. **Mitigation:** Cognitive load optimization in the framework tackles this by deliberately keeping the human engaged (maintaining germane load). The AI should be designed to *prompt human thinking*: asking questions, encouraging reflection, and sometimes even introducing dissonance to ensure the human doesn't just default to acceptance. For instance, the AI could present multiple alternatives and ask the user to choose (thus the user must think and justify their choice). Another mitigation is to integrate *verification prompts*: after the AI gives an answer, it might follow up with "Does this answer make sense to you? Why or why not?" to nudge the user to evaluate it. In educational settings, open-book exam approaches can be instructive – let the AI give information, but then test the learner in a context where they must recall or apply it without AI. If the framework is implemented in a learning platform, it could periodically switch into a mode where AI help is unavailable, simulating real conditions (e.g., an exam or real-life scenario), so users practice performing solo. This not only assesses their learning, it also reinforces that the AI is a tool to learn *with* but not to lean *on* unthinkingly. Designing AI explanations that are a bit incomplete can also be useful; for example, the AI might give a solution but not explain one step, leaving a "gap" for the learner to fill (similar to how a good textbook leaves some proofs as exercises for the reader). By ensuring the human always has an active role, we mitigate the risk of cognitive complacency and keep their skills sharp.

- **Bias Amplification and Cognitive Distortions** – AI systems can carry biases from their training data or programming, and a human who learns from or collaborates with such an AI might quickly adopt those biases, potentially even more rapidly than through human peers. Glickman & Sharot [21] noted that because AI is often perceived as knowledgeable and provides a high signal-to-noise ratio, people tend to accept its biases faster, leading to a phenomenon of "hyper-learning" of biases. This is a form of cognitive distortion, where one's views may skew toward the AI's outputs (which could systematically underrepresent certain perspectives or overrepresent particular patterns). In a feedback loop, humans interacting with biased AI can become more biased themselves, which in turn could feed back into AI (if the AI learns from user interactions). Beyond social biases, there are also distortions like *confirmation bias* and *echo chambers*: if an AI tailors itself too much to user preferences (personalization gone wrong), it might only show information that confirms the user's beliefs, limiting exposure to diverse viewpoints. **Mitigation:** The framework's adaptive personalization must be coupled with diversity and fairness considerations. One approach is to ensure AI scaffolds present *alternative viewpoints or solutions* deliberately, to counteract any single biased narrative. For example, an AI writing assistant might flag when a user's essay lacks counterarguments and suggest considering the opposing side. In learning environments, the AI can be programmed to avoid reinforcing a student's misconceptions; instead of just giving the correct answer which the student might accept blindly, it can prompt the student to reason through their mistake. Additionally, transparency about uncertainty or bias in AI outputs is important. If the AI is unsure or the domain is contentious, it should communicate that (e.g., "There are multiple opinions on this issue…"). This invites the human to engage in critical evaluation rather than take the AI's word as gospel. From a design perspective, using diverse training data and bias mitigation techniques in the AI itself is essential to minimize harmful biases from the start. But since no AI is bias-free, the scaffolding framework could include a meta-cognitive layer: training users in AI literacy, teaching them that AI can be wrong or biased. In essence, the scaffold should include a gentle warning scaffold: reminding users to verify from other sources or encouraging them to be skeptical when appropriate. By treating human and AI as collaborative partners, the framework can incorporate *bias checking as a joint responsibility*. For instance, an AI tutor could occasionally ask a student, "How could we double-check this information?" to build the habit of verification.

- **Reduction in Diversity of Thought and Creativity** – If many people rely on similar AI tools, there is a concern of *homogenization* of outputs. Ivcevic & Grandinetti [22] and others [26] observed that while individuals might become more productive with AI, the collective diversity of ideas can shrink because the AI often leads users down similar paths. This is a "*social dilemma*" where everyone individually benefits (e.g., faster generation of ideas), but if the AI's style or suggestions have common patterns, the variety of independent thought in a group or society could diminish. In creative industries, this could mean a thousand designs start to look the same because they all began from the same AI templates. In learning, students might converge on similar essay answers if they all use the same AI hints. In problem-solving, humans might neglect out-of-the-box approaches that the AI didn't suggest, leading to a narrowing of exploration. **Mitigation:** To preserve and indeed encourage diversity of thought, the scaffolded AI should avoid being too prescriptive or convergent. Adaptive personalization can help here by varying the support per individual – if each person's AI experience is a bit different, their outcomes will also differ. Another strategy is building in randomness or creativity prompts: the AI might occasionally throw an unconventional prompt or encourage the user to find a second solution different from the first. In collaborative settings, different AI instances could play devil's advocate to each other: imagine a scenario in a classroom where each student's AI tutor gives slightly different perspectives on a debate topic, and then the class can pool those perspectives, yielding a richer discussion than if all had the identical argument. Moreover, the progressive autonomy approach encourages users to add their personal touch as AI support fades. In creativity, for example, once the AI gives an initial idea, the user is prompted to diverge from it or elaborate in a unique direction as the session continues – ensuring the final creation carries the human's individual imprint. We can also mitigate homogenization by exposing users to *multiple AIs or tools* – e.g., not relying on a single model for everything, which can reduce single-source bias. Finally, fostering an attitude of creativity in users themselves is key: the AI scaffold can include motivational prompts like "What's a different angle you can think of?" or "Try to surprise me with a solution that I (the AI) wouldn't predict." This turns the potential homogeneity problem into a game where the human actively tries to outthink or extend beyond the AI, thereby maintaining originality and breadth of thinking.

- **Mismanagement of Cognitive Load** – If the cognitive load dimension is not finely tuned, there are two extremes of risk: *overload* and *under-stimulation*. Overload can happen if the AI fails to simplify the task enough or provides *too much information* (information overload). A learner might be overwhelmed with AI-generated content, hints, analytics, etc., to the point of confusion (imagine an AI tutor that dumps five different ways to solve a problem on a student at once – that might increase extraneous load and discourage the student). Under-stimulation, on the other hand, occurs if the AI makes things too easy or intervenes too quickly, leading to boredom or superficial engagement. If the scaffold is too helpful, the user might sail through without truly processing anything (the "easy mode" problem). **Mitigation:**

The adaptive nature of the framework is critical – continuous monitoring of the user's cognitive state (through performance and possibly affect measures) allows the AI to adjust the level of challenge. For overload, the AI should apply the *KISS principle* ("keep it simple, stupid") for extraneous elements: only give the necessary info, perhaps one hint at a time, and gauge the user's response before adding more. Techniques like *adaptive hinting* [19] (where the AI starts minimal and only elaborates if needed) directly prevent overload. For under-stimulation, the AI can introduce desirable difficulties: intentionally not intervening immediately, or adding a twist to the task. Some intelligent tutors implement a "*hint delay*" to give students a chance to think before showing a hint. The scaffold could also escalate the challenge if it detects sustained high performance – for instance, shortening the time limits, introducing open-ended tasks, or even encouraging the user to teach back the material (teaching is a high-germane-load activity that solidifies learning). By dynamically keeping the task in a challenging-but-doable range, the framework mitigates both extremes. Additionally, obtaining user feedback explicitly ("Was that too hard or too easy?") can help calibrate difficulty. In collaborative environments, humans can also help each other – the AI might pair up users to explain things to one another, which can re-engage someone who was under-challenged by making them the explainer, or reduce overload by having peers share strategies. The scaffold thus doesn't operate in isolation; it can leverage social learning to regulate cognitive load as well.

- **Ethical and Trust Concerns** – Lastly, a broader risk is if the AI scaffold is not trusted or is used inappropriately. If users suspect that reliance on the AI is making them less skilled, they may resist using it (possibly a justified fear if they've seen others become too dependent). Conversely, if they trust it too much, they might follow it even when it's wrong (as noted in clinical settings where AI suggestions can sometimes be accepted even when incorrect, leading to errors [41]). There's also the risk of privacy and surveillance [42] if the AI is constantly monitoring performance to personalize – users might feel uncomfortable or judged.

**Mitigation:** Building trust through explainability and user control is important [43]. The AI should be able to explain its suggestions in a way that the user can learn ("I suggested this because…"), which also helps the user know when it might be off. In critical domains like healthcare or finance, keeping a human-in-the-loop for final decisions is essential – the scaffold is an advisor, not an autonomous actor, which should be made clear. Users should be encouraged to question the AI and not feel that that's "going against" the system. Culturally, organizations and educators can emphasize that AI is a tool for enhancement, not a cheat or a replacement. By aligning the AI's role with human values (learning, growth, safety), and allowing opt-outs or custom settings, users are more likely to embrace the scaffold. The framework's aim to promote autonomy itself helps here: users see that the AI is making them better, not obsolete. If implemented well, over time users trust the scaffold because they feel their own competence rising, which validates the process. Finally, ethics training around AI use (for both developers and users) can foresee and address issues like fairness, transparency, and the importance of maintaining one's skills, thereby preemptively mitigating misuse.

## 4.1 Framework Safeguards

Each dimension of the enhanced scaffolding framework inherently contains *safeguards* against these risks:

- **Progressive autonomy** counters overdependence by design: the AI will not continue to do for the user what the user can eventually do themselves. It ensures a transfer of responsibility. If a user is hesitant to take over, the scaffold can enforce it gently (for example, by saying "Now you try the next one on your own – I'll be here if needed"). This builds confidence and reduces long-term dependency.
- **Adaptive personalization** not only boosts effectiveness but also can serve as a safety mechanism: it can personalize not just for skill, but for *behaviors*. If a user is showing signs of misuse (e.g., always asking the AI for answers without trying), the AI can detect this pattern and adapt by giving more Socratic help instead of direct answers. If a user is disengaging, it can re-engage them with a new approach. Personalization means problems like complacency or frustration can be noticed early and addressed on an individual basis.

- **Cognitive load optimization** is in itself a balance that prevents overload (which can cause users to abandon the tool or learn incorrectly under stress) and underload (which leads to laziness and shallow learning). By keeping the cognitive challenge optimal, it inherently prevents some of the misuse – when tasks are neither too trivial nor impossibly hard, users have the best chance to stay diligent and think critically. Also, part of cognitive optimization is encouraging reflection, which is a built-in antidote to blindly following AI or absorbing biases. If the scaffold routinely asks "why" and "how do you know," the user is habituated to justify and critique, making them less likely to accept errors or biases without noticing.

Furthermore, a crucial element in mitigation is *user education [44] and meta-cognitive awareness [45]*. The enhanced scaffolding framework isn't just about doing things to the user; it can also *teach the user about itself*. For example, an AI tutor might include a brief orientation: "Here's how I will help you: I'll give you hints and gradually give you more independence. If you find yourself unsure, it's okay – that's part of learning. And remember to always think about my suggestions and whether they make sense to you." A workplace AI assistant might similarly be introduced with guidelines: "Use me to assist your work, but make sure you review what I produce." By being explicit, we empower users to be partners in the scaffolding process, not passive consumers of AI help.

Some recent frameworks in AI ethics also stress *human-AI teaming* principles that align well with enhanced scaffolding – for instance, keeping the human authority, ensuring AI augments rather than diminishes human skills, and maintaining vigilance for feedback loops that can degrade outcomes [14]. The scaffolding paradigm naturally incorporates those, since the end goal is a skilled, autonomous human, not a subservient one.

In conclusion, while there are genuine risks of dependency, bias, and cognitive pitfalls, the Enhanced Cognitive Scaffolding framework actively addresses these through its core design. By fading support, it prevents long-term dependency; by adapting, it catches and corrects issues in usage; and by optimizing cognitive engagement, it keeps the human mind firmly in the loop. The result is a self-correcting system that not only scaffolds the task at hand but also scaffolds the user's relationship with AI itself – guiding them toward a healthy, critical, and ultimately independent interaction with technology.

## 5. Conclusion

AI systems, when thoughtfully designed, can function as powerful cognitive scaffolds that amplify human learning and performance while safeguarding against the pitfalls of automation. The Enhanced Cognitive Scaffolding framework we have outlined embodies this philosophy, operationalizing Vygotskian guided learning in the age of AI. By integrating progressive autonomy, adaptive personalization, and cognitive load optimization, AI becomes a personalized cognitive partner – initially an expert guide and eventually a supportive peer – that helps users climb to new levels of competence and then steps aside so they can stand on their own. We illustrated how this works in practice across education, workplace training, creative endeavors, and healthcare: in each case, the AI provides high support at the outset (whether through detailed hints, tailored coaching, or information management) and then gradually cedes control to the human as they grow more capable. This process not only accelerates skill acquisition and problem-solving success, but does so in a way that empowers the human – building confidence, independence, and mastery.

Crucially, this framework is not naive about the challenges. We have seen that without careful design, AI assistance can lead to dependency, reduced critical thinking, bias reinforcement, and homogenization of thought. However, the enhanced scaffolding approach contains the remedies to its own risks: it tapers off assistance to force independent thinking (addressing dependency), keeps the user mentally active (preventing complacency), and adapts to user behavior (mitigating bias and engagement issues) to ensure AI is a *help* not a *hinderance*. As Wang & Fan [17] emphasize, AI tools like ChatGPT yield the best educational outcomes when embedded in a proper scaffolded framework that guides their use. In other words, technology alone is not a silver bullet – it's how we structure the human-AI interaction that determines whether we get amplification of learning or erosion of it. The enhanced scaffolding framework provides that structure, marrying proven pedagogical strategies with the dynamic capabilities of AI.

Looking ahead, the implications of widespread AI cognitive scaffolding are profound. Education could become more learner-centric than ever, with each student receiving a personal AI mentor that not only teaches but learns how to teach *them* best. Workplaces could see continuous upskilling, where employees effectively have an on-demand coach for any new task, ensuring that as jobs evolve, workers evolve in tandem rather than being left behind. Creative collaboration between humans and AIs could push the boundaries of innovation, with AI scaffolds encouraging humans to explore ideas they might never have attempted solo, yet keeping the human creative process at the heart of it. In healthcare and other high-stakes fields, AI scaffolding could enhance decision-making and training while keeping human judgment in control, potentially leading to better outcomes and greater safety. Importantly, because the focus is on progressive autonomy, the end result is *not* a world where humans are coddled by machines, but rather one where humans are elevated by machines – equipped with greater knowledge, sharpened skills, and the ability to tackle challenges that were previously beyond their reach.

In sum, AI as a cognitive scaffold represents a shift from viewing AI as an autonomous problem-solver to viewing it as an *enabler of human growth*. It recalls the old proverb: "Give a man a fish and you feed him for a day; teach a man to fish and you feed him for a lifetime." Traditional AI might "give the answers" (the fish), but enhanced cognitive scaffolding aims to "teach the skills" (fishing). By fostering a symbiotic learning partnership between human and AI, we can achieve learning and performance outcomes neither could accomplish alone – all while ensuring the human emerges more competent and autonomous. This synergy of artificial and human intelligence, carefully balanced, could be one of the most positive developments for education and work in the coming years. The task now for researchers, designers, and educators is to continue refining this framework, testing it in diverse real-world scenarios, and updating it with ethical guardrails, so that the vision of AI as a nurturing cognitive scaffold becomes a widespread reality rather than an exception. The evidence so far is encouraging, and with ongoing interdisciplinary effort, enhanced cognitive scaffolding could truly transform AI from a potential threat to an indispensable ally in human cognitive development [18, 21].

Ultimately, the success of this approach will be measured by a simple outcome: *learners and workers who engage with AI scaffolds should become better at what they do – more knowledgeable, more skilled, and more independent – than they were before*. If we achieve that on a broad scale, it will validate the promise that AI, rather than making humans obsolete, can help each of us reach our fullest potential.